\documentclass{article}

\usepackage{arxiv}

\usepackage[utf8]{inputenc} 
\usepackage[T1]{fontenc}    
\usepackage{hyperref}       
\usepackage{url}            
\usepackage{booktabs}       
\usepackage{amsfonts}       
\usepackage{nicefrac}       
\usepackage{microtype}      
\usepackage{lipsum}
\usepackage{graphicx}
\graphicspath{ {./images/} }
\usepackage{mathtools}
\usepackage{multirow} 
\usepackage[noadjust]{cite}

\begin{document}

\title{Deep learning in physics: a study of dielectric quasi-cubic particles in a uniform electric field} 

\author{
  Zhe Wang\\
  Energy Research Institute\\ 
  Nanyang Technological University\\
  637141 Singapore \\
  \texttt{zhe.wang@ntu.edu.sg} \\
   \And
  Claude Guet \\
  Energy Research Institute\\
  Nanyang Technological University\\
  637141 Singapore \\
  School of Materials Science and Engineering \\
  Nanyang Technological University \\
  639798 Singapore\\
  \texttt{cguet@ntu.edu.sg} \\
}

\maketitle

\begin{abstract}
Solving physics problems for which we know the equations, boundary conditions and symmetries can be done by deep learning. The constraints can be either imposed as terms in a loss function or used to formulate a neural ansatz. In the present case study, we calculate the induced field inside and outside a dielectric cube placed in a uniform electric field, wherein the dielectric mismatch at edges and corners of the cube makes accurate calculations numerically challenging. The electric potential is expressed as an ansatz incorporating neural networks with known leading order behaviors and symmetries and the Laplace's equation is then solved with boundary conditions at the dielectric interface by minimizing a loss function. The loss function ensures that both Laplace's equation and boundary conditions are satisfied everywhere inside a large solution domain. We study how the electric potential inside and outside a quasi-cubic particle evolves through a sequence of shapes from a sphere to a cube. The neural network being differentiable, it is straightforward to calculate the electric field over the whole domain, the induced surface charge distribution and the polarizability. The neural network being retentive, one can efficiently follow how the field changes upon particle's shape or dielectric constant by iterating from any previously converged solution. The present work's objective is two-fold, first to show how an a priori knowledge can be incorporated into neural networks to achieve efficient learning and second to apply the method and study how the induced field and polarizability change when a dielectric particle progressively changes its shape from a sphere to a cube.
\end{abstract}


\section{Introduction}\label{sec:Introduction}

Solving physics problems for which we know the equations, boundary conditions and symmetries by deep learning follows from the universal approximation theorem \cite{Gorban98,Winkler17,Lin18}, which states a sufficiently deep artificial neural network (ANN) can approximate any well-behaved function with a finite number of parameters. In 1994, Meade and Fernandez approximated the solution of linear \cite{Meade94} and nonlinear \cite{Meade94b} ordinary differential equations using a single-layer perceptron. This approach was soon generalized by Lagaris et. al. \cite{Lagaris98} and applied to two-dimensional Poisson equations with various source terms in a rectangular domain. Their trial functions consisted of two terms: the first one which contained no trainable variable satisfied the boundary conditions, whereas the second one involved neural networks, whose parameters, e.g. weights and biases, were adjusted to minimize a loss function. A neural network solution of Poisson's equation in a 3D domain with irregular boundaries was achieved by including the boundary conditions into the loss function \cite{Lagaris00}. Alternatively, McFall and Mahan constructed a proper trial function so that the boundary conditions were automatically satisfied \cite{McFall09}. Massive growth of available scientific data has introduced a new flavor into the ANN approach to differential equations. By incorporating data and governing equations into the loss functions, the physics-informed neural networks enable inferring hidden physics from measured data \cite{Raissi18a,Yang19,Rackauckas20} with successful applications on the visualization of turbulent flows \cite{Raissi18,Raissi19,Raissi20}, where multiple scales interact, and the design of metamaterials in nano-optics \cite{Chen20}, where the finite size effect dominates. However, as pointed out by Wong et al. \cite{Wong21}, high-dimensional, non-convex loss functions require significant optimization efforts, calling for sophisticated hyper-parameter optimization \cite{Wang21a,Wang21b,Elhamod21} and transfer neuroevolution \cite{Wong21} algorithms.

In this work, we combine physics knowledge with recent advances in deep learning to calculate the induced electric field of semiconductor colloidal nanocrystals in an external photon field. Semiconductor colloidal nanocrystals show numerous advantages for powerful opto-electronic materials, as their optical properties change with shape and size due to quantum effects \cite{Klimov,Delerue}. In order to characterize correctly their absorption/emission properties, one needs to know the electric field induced by the external photon field \cite{BeckerNatLett2018,Nguyen20}. As the particle's size is much smaller than the photon wavelength, the inner field results from a homogeneous applied electric field of amplitude given by the laser intensity. Whereas there are analytical solutions of Laplace's equation for dielectric particles with shapes that allow to define a system of curvilinear coordinates such as sphere, ellipsoids, torus, etc. \cite{Klimov,Landau_electrodynamics,Jackson,Stratton_book}, numerical solvers are required for other shapes. The case of a cube is challenging because the edges and corners lead to sharp variations of the induced fields. To our knowledge there is only one published calculation based on finite element methods applied to a cube with relative dielectric constant ranging from to $2$ to $10$ \cite{BeckerNatLett2018}. The authors claimed a $1\%$ accuracy at the far field of their finite element simulation domain. At low values of the relative dielectric constant considered, they found that the electric field at the center of the particle is lower for a cube than for a sphere. This would imply a lower polarizability for a cube than for a sphere, at variance with accurate calculations \cite{Edwards61, Herrick77, Eyges79,Sihvola01,Sihvola04,Sihvola07,Helsing13}. Note that these herein cited accurate calculations all focused on solving a surface integral equation to estimate the dielectric polarizability and consequently did not provide the induced electric field inside and outside the cube.

In this paper, we introduce an alternative method to calculate the electric field inside and outside a dielectric nanoparticle embedded in another dielectric medium. From the field inside, the polarizability is extracted. More specifically, we approximate the solution of Laplace's equation by a function combining a known analytical term and an ANN. Then, instead of solving numerically Laplace's equation with boundary conditions, we express the problem as an optimization problem and construct a loss function which can be minimized by machine learning algorithms to yield the full electric potential inside and outside the particle. One clear advantage of the ANN approach over finite element methods is that it provides the solution of Laplace's equation as a differentiable function that can be used as such. Additionally, the retentive nature of neural networks allows for a systematic tracking of the evolution of induced fields as dielectric particles deviate from canonical spherical and cubic shapes. Note that physical nanoparticles usually have rounded edges and corners, see e.g. Fig.~2s in \cite{Kovalenko17} and in \cite{Tang20} for perovskite nanocrystals.

Most previous works considered partial differential equations in a homogeneous domain with explicit boundary conditions. The electrostatic problem of a colloidal particle requires to treat accurately the discontinuity of the displacement field at the interface. To the best of our knowledge, the present work is the first ANN-based attempt to address this problem, namely solving Laplace's equation in a three-dimensional piece-wise homogeneous domain with implicit boundary conditions on an irregular interface. In addition to formulating the best loss functions, one can substantially reduce the optimization effort by incorporating known symmetries and specific features, e.g. leading order behaviors, into the ANN ansatz.

The rest of the paper is organized as follows: in Sec. \ref{sec: math}, we present the physics model. Starting from a general solution in the form of a linear combination of spherical harmonics, we replace the higher order terms by a function containing neural networks and construct a loss function that includes all constraints. The architecture of the neural network is then discussed. In Sec. \ref{sec: application-spheroid} , we benchmark the proposed ANN method by studying dielectric spheroids for which analytical solutions are known \cite{Landau_electrodynamics}. In Sec. \ref{sec: application-quai-cube}, we study the evolution of polarizabilities and induced fields inside and outside a dielectric particle through a sequence of shapes from a sphere to a cube. Finally, conclusions are drawn in Sec. \ref{sec: conclusion}, with a highlight on future works.

\section{ Neural Network solution of Laplace's equation }  \label{sec: math}

\subsection{Governing equations}

Consider a neutral homogeneous dielectric 3D particle, with surface $S$ and permittivity $\epsilon_1$, embedded in a homogeneous medium with permittivity $\epsilon_0$. In a spherical coordinates  $(r, \theta, \varphi)$ system whose origin coincides with the center of the particle, we take the direction of an uniform electric field $\boldsymbol{E}_{\text{ext}}$ to be the axis from which the polar angle $\theta$ is measured. In the absence of external charges, the electric potential obeys Laplace's equation inside and outside the particle,
\begin{align} \label{eqn 1}
\nabla^2 \phi  = 0. 
\end{align}
The continuity of the tangential components of the electric field ($\boldsymbol{E} = -\nabla\phi$) and the normal component of the displacement field ($\boldsymbol{D} = \epsilon\boldsymbol{E}$) at the interface $S$ leads to the following boundary conditions for the potential:
\begin{subequations} \label{eqn 2}
\begin{align}
    \phi_{0} &= \phi_{1},  \label{eqn 2a} \\ 
    \nabla \phi_{0} \cdot \hat{\boldsymbol{n}} &= \epsilon_r \nabla \phi_{1} \cdot \hat{\boldsymbol{n}}. \label{eqn 2b}
\end{align}
\end{subequations}
Here, $\hat{\boldsymbol{n}}= \nabla S/ \vert \nabla S\vert$, is the unit normal vector to $S$, $\epsilon_r = \epsilon_1/\epsilon_0$ is the relative dielectric constant, and subscripts $_{0}$ and $_{1}$ denote quantities outside and inside the particle, respectively. Furthermore, the electric field tends asymptotically to the applied field and the potential is zero at the origin, yielding
\begin{subequations} \label{eqn 3}
\begin{align}
\phi_{0} &= -E_{\text{ext}} r \cos(\theta), \quad \mbox{as} \quad r \to \infty, \\
\phi_{1} &= 0, \quad \mbox{at} \quad r = 0,
\end{align}
\end{subequations}
where $E_{\text{ext}} = \vert \boldsymbol{E}_{\text{ext}} \vert$.

\begin{figure}[b]
\centering
\includegraphics[width=0.75\textwidth]{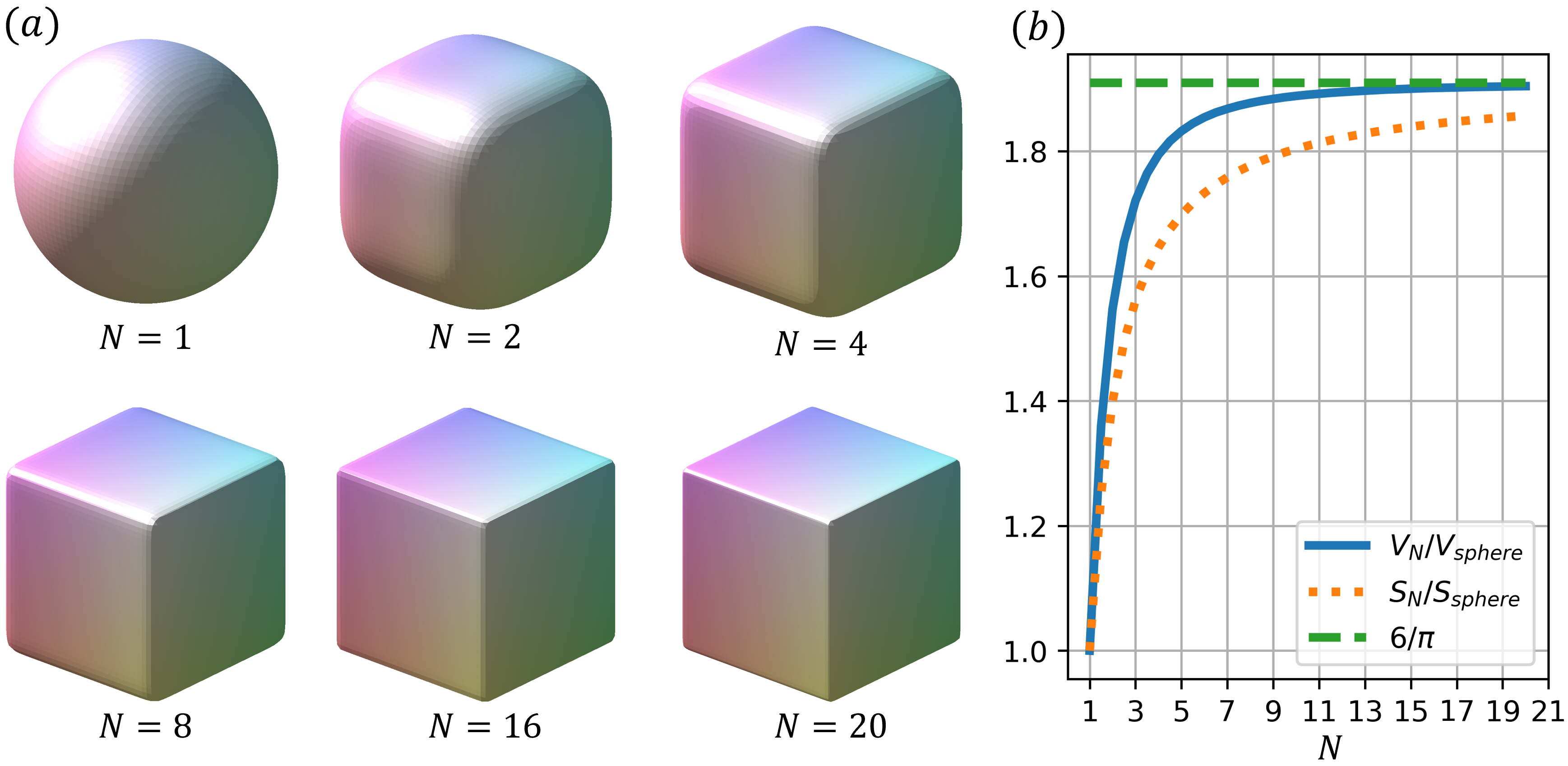}
\caption{\label{fig:super_ellipsoid} $(a)$ Sequences of shapes from a sphere to a quasi-cube with increasing values of $N$, cf. Eq.~(\ref{eqn 4}) with $a=b=c$. $(b)$ Plot of normalized volume and surface area of quasi-cubes by the same of the sphere as a function of $N$. }
\end{figure}

Within the perspective of approaching a real cube and benchmarking the present numerical method with analytical solutions, we consider a super-ellipsoidal inclusion whose surface is described by the equation
\begin{align} \label{eqn 4}
    S(x, y, z) \equiv \left\vert\frac{x}{a}\right\vert^{2N} + \left\vert\frac{y}{b}\right\vert^{2N} + \left\vert\frac{z}{c}\right\vert^{2N} = R^{2N},
\end{align}
in the Cartesian coordinates. Here, $a, b, c, R \in \mathbb{R}^{+}$ and the exponent $N \geq 1$. For $a=b=c$ and $N=1$, Eq.~(\ref{eqn 4}) defines a sphere with radius $R$. A continuous deformation of a sphere with radius $R$ to a cube of edge length $2R$ as $N$ increases towards $N \in \infty$, is shown in Fig.~\ref{fig:super_ellipsoid}. For cases where $a$, $b$, $c$ are not equal to each other, one obtains an ellipsoid for $N=1$ and a rectangular parallelopiped as $N\to\infty$. Note by passing that ``real'' nanoparticles have rounded corners, resembling quasi-cubes with $N \in [1, 8]$ in Fig.~\ref{fig:super_ellipsoid}. For instance, by fitting $2D$ images of $100$ perovskite nanocrystals to a superellipse, cf. Eq.~(\ref{eqn 4}) with $z=0$, Tang et al. \cite{Tang20} found that $N \approx 2.65$ for freshly synthesized and $N \approx 1.8$ for aged samples, respectively. 

A general solution to Eq.~(\ref{eqn 1}) can be expressed as a linear combination of spherical harmonics $Y_l^m(\theta, \varphi)$ weighted by appropriate scaling factors $r^{l}$ inside and $r^{-(l+1)}$ outside of the dielectric particle
\begin{equation} \label{eqn 5}
    \phi = \sum_{l=0}^{\infty} \sum_{m=-l}^{l} \left[ A_l^m r^{-(l+1)} + B_l^m r^l \right] Y_l^m(\theta, \varphi),
\end{equation}
where coefficients $A_l^m$ and $B_l^m$ are determined by the boundary conditions. The homogeneity and the direction of the external electric field $\boldsymbol{E}_{\text{ext}}$ along the $z$-axis impose an anti-symmetry on the electric potential with respect to the mid-plane $\theta = \pi/2$, which is orthogonal to $\boldsymbol{E}_{\text{ext}}$ and crosses the center of the particle. Thus,
\begin{align} \label{eqn 12}
    \phi(r, \theta, \varphi) = -\phi(r, \pi-\theta, \varphi),
\end{align}
such that only odd terms remain in the spherical harmonics expansions (\ref{eqn 5}). For sake of simplicity, we select one of the principal axis of the dielectric particle to be aligned with $\boldsymbol{E}_{\text{ext}}$. As such, there is a further mirror symmetry,
\begin{subequations} \label{eqn symmetry_varphi}
\begin{align}
    \phi(r, \theta, \varphi) &= \phi(r, \theta, - \varphi), \\
    \phi(r, \theta, \varphi) &= \phi(r, \theta, \pi + \varphi), \\
    \phi(r, \theta, \varphi) &= \phi(r, \theta, \pi - \varphi).
\end{align}
\end{subequations}

For $N=1$, only the dipole moment contributes to $\phi_{1}$, leading to a constant electric field inside the particle \cite{Landau_electrodynamics}. Deviating from a sphere, $N \geq 2$, however, brings in contributions from higher-order moments with no clear optimal order of truncation, rendering an analytical solution unfeasible.

\begin{figure*}
   \centering
   \noindent\makebox[0.95\textwidth]{%
   \includegraphics[width=\textwidth]{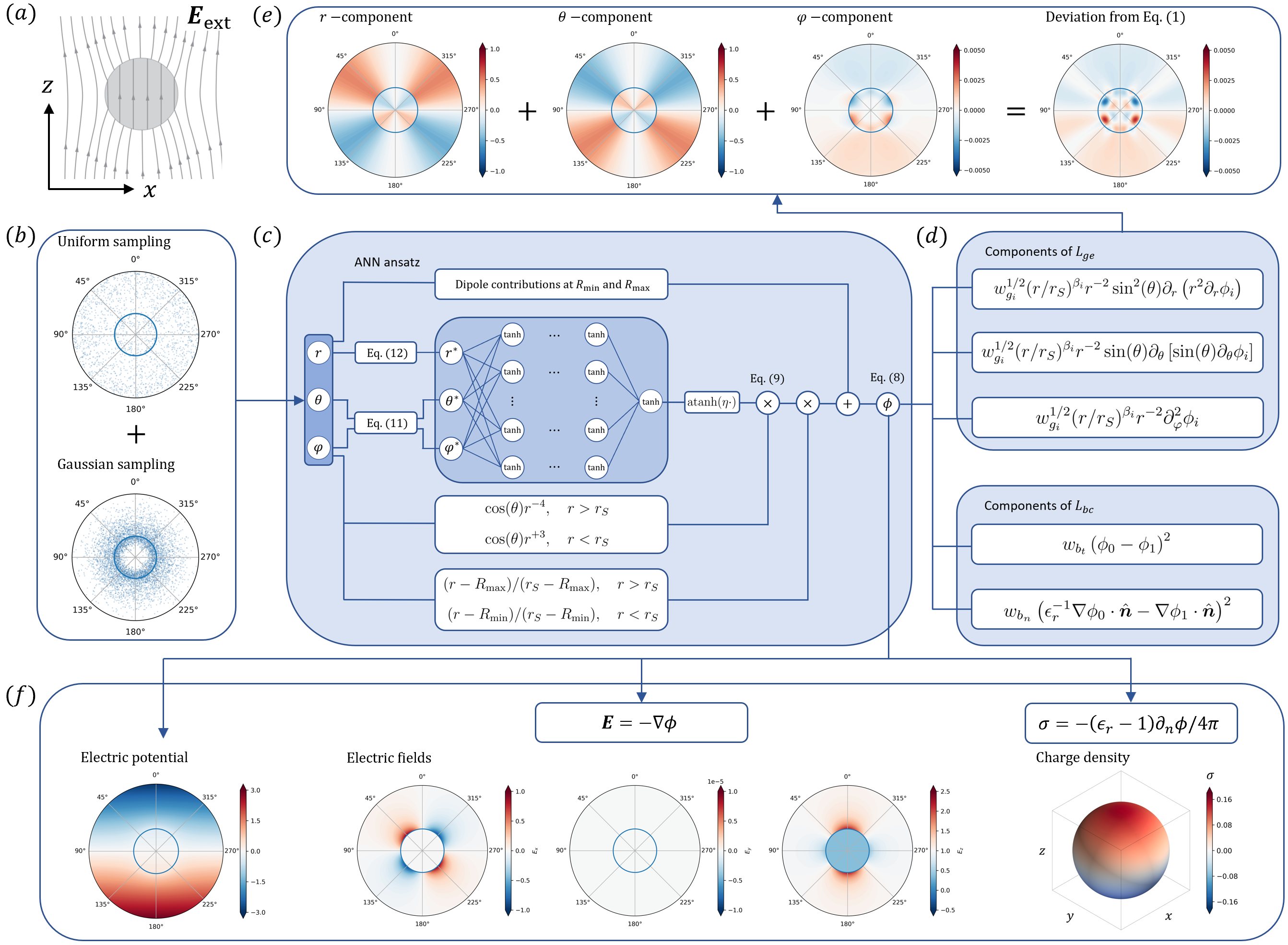}}
   \caption{\label{fig:flowchart} $(a)$ Sketch for a dielectric particle immersed in a uniform electric field along the $z$-axis, $\boldsymbol{E}_\text{ext}$. $(b)$ A superposition of collocation points sampled from a uniform distribution and a Gaussian distribution with mean centered at the dielectric interface $r_S$. $(c)$ Proposed ANN ansatz for electric potentials, see Sec.~\ref{sec: ANN_ansatz}. $(d)$ Loss function which measures the deviation from the Laplace equation (\ref{eqn 1}) and the boundary conditions (\ref{eqn 2}) on the dielectric interface, see Sec.~\ref{sec: loss}. $(e)$ Visualization of radial, polar, and azimuthal components of $\nabla^2 \phi$ and its deviation from Eq.~(\ref{eqn 1}). $(f)$ Recovery of electric fields $\boldsymbol{E}$ and charge density $\sigma$ from $\phi$. Here and thereafter, results are plotted in a circular domain of radius $3$, wherein the blue curve indicates the interface.}
\end{figure*}

\subsection{ANN ansatz} \label{sec: ANN_ansatz}

Our ansatz for solving Laplace's equation with the boundary conditions given by Eqs.~(\ref{eqn 2},\ref{eqn 3}) is to explicitly retain the dipole contributions and group the infinite series of higher order terms into two ANN functions:
\begin{subequations} \label{eqn aphi}
\begin{align}
    \phi_{0} &= -\left[E_{\text{ext}} r - d_0 r^{-2} \right] \cos(\theta) + \frac{r - R_{\text{max}}}{r_S - R_{\text{max}}} H_0, \label{eqn aphi0}\\
    \phi_{1} &= d_1 r \cos(\theta) + \frac{r - R_{\text{min}}}{r_S - R_{\text{min}}} H_1, \label{eqn aphi1}
\end{align}
\end{subequations}
where the dipole coefficients $d_0$ and $d_1$ are unknown. Here, $r_S = r_S(\theta, \varphi)$ is an implicit solution to Eq.~(\ref{eqn 4}) measuring the distance from the origin to points on the interface, $R_{\text{min}}$ and $R_{\text{max}}$ are the inner and outer boundaries of the solution domain and finally, $H_0$ and $H_1$ denote ANN based functions modeling a deviation from the leading order dipolar behavior of the induced electric field. A breakdown of the proposed ansatz (\ref{eqn aphi}) is detailed below.

To get around the singularity at the origin of the Laplacian operator in spherical coordinates \cite{Khelashvili15}, we assume a practical origin of the radial coordinate: $R_{\text{min}} \ll 1$. Similarly, to enable a numerical calculation, the infinity $r \to \infty$ is replaced by a practical infinity: $R_{\text{max}} \gg 1$. To mitigate the effect of such a truncation, we modify the boundary conditions (\ref{eqn 3}) by including the dipole contributions at the practical infinity and practical origin, leading to the first terms on the right hand side of the ANN ansatz (\ref{eqn aphi}). Then, the approximate solution domain is a spherical shell with inner and outer radii $R_\text{min}$ and $R_\text{max}$, wherein the modified boundary conditions are satisfied by construction. In the limit, $R_{\text{min}} = 0$ and $R_{\text{max}} \to \infty$, boundary conditions (\ref{eqn 3}) are recovered.  

To inform ANN ansatz (\ref{eqn aphi}) with the leading octupolar radial trends, namely $r^{+3}$ inside and $r^{-4}$ outside the particle, as well as the symmetry constraints of Eqs.~(\ref{eqn 12},\ref{eqn symmetry_varphi}), we select
\begin{subequations} \label{eqn 10}
\begin{align} 
    H_0 &= \cos(\theta) r^{-4} \text{atanh}\left[\eta \text{NN}_0(r^*, \theta^*, \varphi^*; \boldsymbol{\xi}_0) \right], \\
    H_1 &= \cos(\theta) r^{+3} \text{atanh}\left[\eta \text{NN}_1(r^*, \theta^*, \varphi^*; \boldsymbol{\xi}_1) \right],
\end{align}
\end{subequations}
so that the parameters $\boldsymbol{\xi}_i$ of neural networks $\text{NN}_i$, with $i = 0, 1$, are adapted to learn deviations from octupolar radial trends. The symmetry constraints are imposed on $\text{NN}_i$ through a reparameterization of spatial variables $(r^*, \theta^*, \varphi^*)$ and the inclusion of $\cos(\theta)$. The neural network architectures, the constant $\eta$, and the reparameterized coordinate variables $(r^*, \theta^*, \varphi^*)$, are discussed below.

We describe both $\text{NN}_{i}$ as multilayer perceptrons \cite{Cybenko89} each consisting of four densely connected hidden layers with $16$ neurons per layer
\begin{align} \label{eqn mlp}
    \boldsymbol{x}^{[k]} = \tanh( \boldsymbol{W}^{[k]} \cdot \boldsymbol{x}^{[k-1]} + \boldsymbol{b}^{[k]} ),
\end{align}
where $k$ is the index of the current layer, and the weight $\boldsymbol{W}^{[k]}$ and bias $\boldsymbol{b}^{[k]}$ operate a linear transformation of the input vector $\boldsymbol{x}^{[k-1]}$. As the bias of both input and output layer are selected to be zero vectors, the neural network eventually consists of $880$ trainable variables. Instead of the canonical rectified linear unit (ReLU) which vanishes upon second-order differentiation, we assign the activation function to be the hyperbolic tangent function. Since the output of $\text{NN}_i$ measures a deviation from the octupolar trend, it must remain bounded, with an amplitude that depends on the strength of the external field, the geometry of the dielectric inclusion and the mismatch at the dielectric interface. Therefore, an additional operation $\text{atanh}(\eta \cdot)$ is included in Eqs. (\ref{eqn 10}) to transform the output of $\text{NN}_i$, ranging from $[-1,1]$, into an interval $[-\text{atanh}(\eta), \text{atanh}(\eta)]$. A comparison of the proposed activation function with the canonical linear and $\tanh$ activation is discussed in Appendix A. In this work, with $E_{\text{ext}} = 1$, we select $\eta = 0.99$. The model parameters $\boldsymbol{\xi}_i = [\boldsymbol{W}_i, \boldsymbol{b}_i]$, as well as $d_i$ defined in Eqs.~(\ref{eqn aphi0}, \ref{eqn aphi1}), are determined by minimizing the loss function discussed in \S \ref{sec: loss}. 

Noting that the Laplacian operator is of second order, we consider the following continuous and differentiable transformation of angular variables
\begin{align}
    \theta^* = -\cos(2\theta) \quad \mbox{and} \quad \varphi^* = -\cos(2\varphi),
\end{align}
which simultaneously maps $\theta \in [0, \pi]$ and $\varphi \in [-\pi, \pi]$ to the first quadrant and rescales them to the range $[-1, 1]$. Similarly, the radial variable $r \in [R_{\text{min}}, R_{\text{max}}]$ is normalized to the range $[-1, 1]$ using the min-max normalization
\begin{eqnarray}
    r^* = \left\{ 
    \begin{array}{c}
    -1 + 2\dfrac{r - \text{min}(r_S)}{R_{\text{max}} - \text{min}(r_S)}, \quad \mbox{for} \quad r > r_S, \\[10pt]
    -1 + 2\dfrac{r - R_{\text{min}}}{\text{max}(r_S) - R_{\text{min}}}, \quad \mbox{for} \quad r < r_S.
    \end{array}
    \right.
\end{eqnarray}
With $(r^*, \theta^*, \varphi^*)$, the outputs of neural networks are symmetric with respect to the mid-plane $\theta = 0$, and the anti-symmetry is enforced by the multipliers $\cos(\theta)$ in Eqs.~(\ref{eqn 10}). A flowchart for the proposed ANN ansatz is sketched in Fig.~\ref{fig:flowchart}$(c)$

\subsection{Loss function} \label{sec: loss}

Given governing equation (\ref{eqn 1}) and boundary conditions (\ref{eqn 2}), we write the loss function as:
\begin{align} \label{eqn loss}
    L = L_{ge} + L_{bc}, 
\end{align}
where $L_{ge}$ and $L_{bc}$ measure the mean squared deviations of the ansatz functions (\ref{eqn aphi0}, \ref{eqn aphi1}) from the exact solutions of the governing equation (\ref{eqn 1}) and the dielectric interface boundary conditions (\ref{eqn 2}), respectively. Although an exact minimization of the loss function (\ref{eqn loss}) ensures the uniqueness of the solution to the boundary value problem, an approximation of that solution by neural networks yields a small nonzero loss. The first term writes as: 
\begin{align} \label{eqn L_g}
    L_{ge} &= \frac{w_{g_0}}{N_0 + N_b} \sum_{j=1}^{N_0 + N_b} \left[ \tilde{r}_j^{\beta_0} \sin^2(\theta_j) \nabla^2 \phi_0(\boldsymbol{r}_j) \right]^2 +     \frac{w_{g_1}}{N_1 + N_b} \sum_{j=1}^{N_1 + N_b} \left[ \tilde{r}_j^{\beta_1} \sin^2(\theta_j) \nabla^2 \phi_1(\boldsymbol{r}_j) \right]^2,
\end{align}
where the $\boldsymbol{r}_j = (r_j, \theta_j, \varphi_j)$ denote the $j$-th collocation points sampled from a superposition of a uniform distribution and a Gaussian distribution centered at $r_S$ as shown in Fig.~\ref{fig:flowchart}$(b)$; $N_0$, $N_1$ and $N_b$ denote the number of collocation points outside, inside, and on the interface of the dielectric particle, respectively. The multiplier $\sin^2(\theta_j)$ is included to compensate the singularity of the Laplacian operator at $\theta = 0, \pi$. The Laplacian has vanishing magnitude at large $r$ values, whereas it is diverging near $R_{\text{min}}$. Therefore, the solutions $\phi_i$ are not equally optimized throughout the solution domain. Inspired by van der Meer et al. \cite{van_der_Meer20}, we introduce the scaling factors $\tilde{r}_j^{\beta_i}$, with $\tilde{r}_j = r_j / r_S(\theta_j, \varphi_j)$ defined at each collocation point. We select the exponents $\beta_0 = 4$ and $\beta_1 = 1$, in order to ensure that the radial, polar, and azimuthal components of $L_{ge}$ are of the same order throughout. A breakdown of the Laplacian $\nabla^2 \phi$ componentwise and its visualization are presented in Fig.~\ref{fig:flowchart} $(d)$ and $(e)$. With increasing $N$, larger and larger derivatives associated with sharper and sharper edges and corners cause a strong mismatch among components of $\nabla^2 \phi_i$ near the interface. Consequently, large oscillations of the loss function emerge during the training process, leading to a slow convergence. In order to balance such a mismatch, weights $w_{g_0}$ and $w_{g_1}$ are introduced in Eq.~(\ref{eqn L_g}) in addition to a normalization by $r_S(\theta_i, \varphi_i)$.  

The dielectric interface boundary conditions (\ref{eqn 2}) lead to the following loss function:
\begin{align}
    L_{bc} &= \frac{1}{N_{b}} \sum_{j=1}^{N_b} \left\{ w_{b_t} \left[ \phi_0(\boldsymbol{r}_j) - \phi_1(\boldsymbol{r}_j) \right]^2 + \frac{w_{b_n}}{\vert \nabla S \vert^2} \left[ {\epsilon_r}^{-1} \nabla \phi_0(\boldsymbol{r}_j) \cdot \nabla S - \nabla \phi_1(\boldsymbol{r}_j) \cdot \nabla S  \right]^2 \right\}.
\end{align}
The weights $w_{b_t}$ and $w_{b_n}$ help balance the losses of tangential and normal boundary conditions during the training. Convergence and accuracy are significantly improved when the normalization factor $\vert \nabla S(r, \theta, \varphi) \vert$ is included. 

\section{Computational results} \label{sec: application}

Throughout the experiments, the loss function (\ref{eqn loss}) was evaluated over a sampling of $2^{n}$ collocation points, with $n=13$ on the boundary, $n=14$ inside, and $n=15$ outside of the particle, respectively. The numbers of collocation points were selected to accommodate the GPU memory. The gradients of the loss function with respect to model parameters ($d_i$ and $\boldsymbol{\xi}_i$) were computed using automatic differentiation \cite{Baydin17}; they were subsequently applied to update the model parameters by using the ADAM optimizer \cite{Kingma15} with a starting learning rate of $10^{-3}$. At each $2,000$ iterations, the learning rates were adjusted to be the twice and half of the current loss magnitude for $d_i$ and $NN_i$, respectively. The collocation points were re-sampled after each $10,000$ iterations, suggesting an unsupervised multi-task learning using mini-batch gradient descent with an infinite set of collocation points. Our numerical models were implemented using Python language and TensorFlow backend \cite{tensorflow}. During the training, we keep $w_{g_1} = w_{b_n} = w_{b_t} = 1$ constant, but with the value of $w_{g_0} \in (0,1]$ varies for different cases. The selection of $w_{g_0}$ is detailed in Appendix B. As a reference, on a workstation equipped with two Nvdia GeForce GTX 1080 Ti graphics cards, each iteration takes around $0.1$ second.

\begin{figure}
   \centering
   \noindent\makebox[0.75\textwidth]{%
   \includegraphics[width=0.75\textwidth]{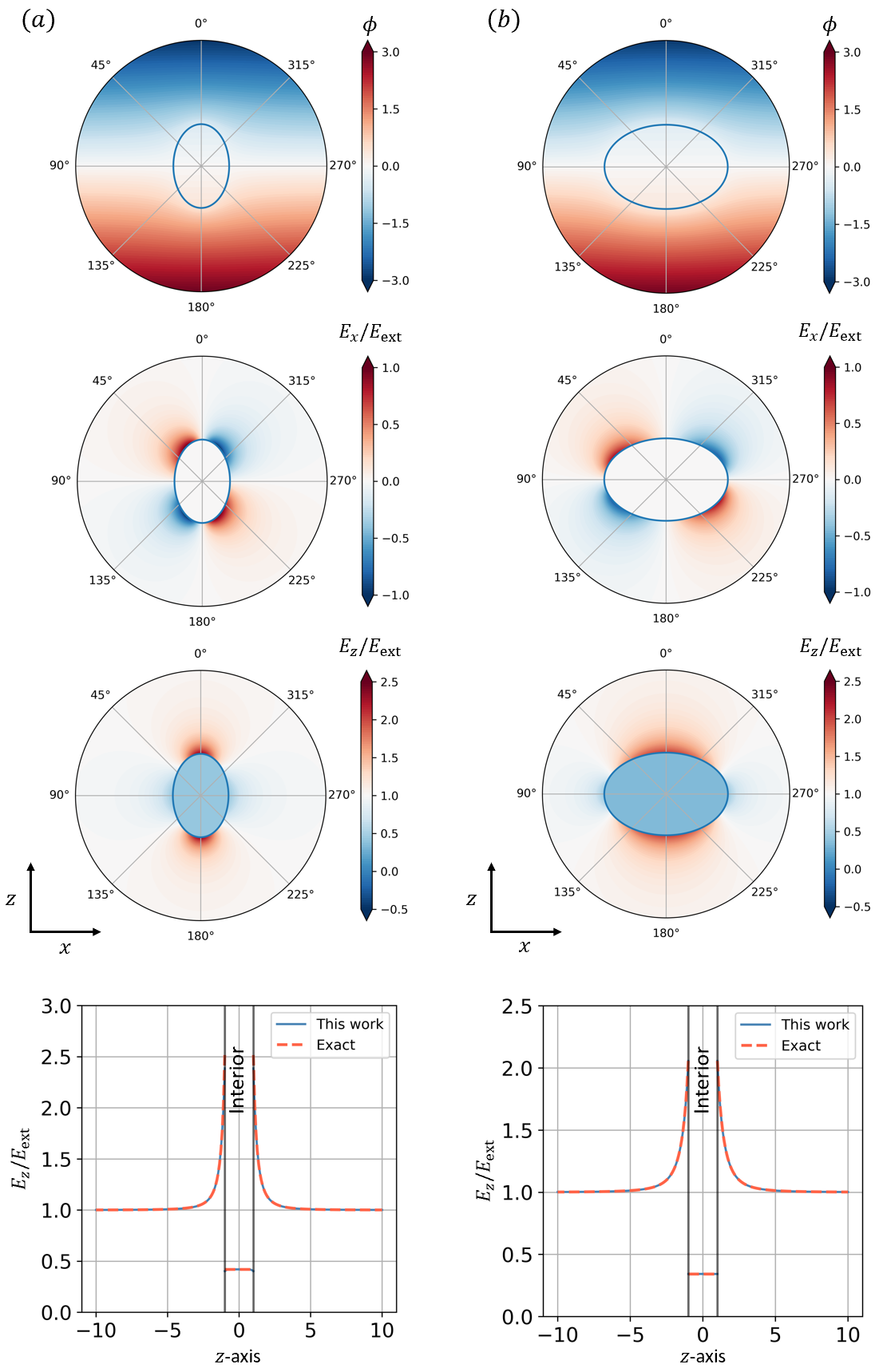}}
   \caption{\label{fig:spheroidal_contour} Dielectric $(a)$ oblate $(a = 2/3, b=c=1)$ and $(b)$ prolate $(a = 3/2, b=c=1)$ particles with $\epsilon_r = 6$ placed in a homogeneous field $\boldsymbol{E}_{\text{ext}} = \hat{\boldsymbol{z}}$. From top to bottom: the electric potential $\phi$, the electric fields $E_x/E_{\text{ext}}$ and $E_z/E_{\text{ext}}$, and a distribution of $E_z/E_\text{ext}$ along principal axes on $xz$-plane spanned by $r$ and $\theta$. As a reminder, the exact electric potential outside a dielectric spheroid ($a, b=c=1$) is: $\phi_0 = -E_0 z \left[ 1 + (\epsilon_r-1) n_z^\xi \right] / \left[ 1 + (\epsilon_r - 1) n_z^\infty \right] $, where $n_z^a = (a/2)\int_{0}^{a} \mathrm{d}s / \left[ (s+1)^2 \sqrt{s + a^2} \right]$ and $\xi = z^2 - 1$ along $z$-axis, cf. \cite{Stratton_book}.}
\end{figure}

The present model allows us to calculate the induced potential over the whole domain. By differentiation, we obtain the total electric field, $\boldsymbol{E}$, the polarization field $\boldsymbol{P}= \frac{\epsilon_r - 1}{4\pi} \boldsymbol{E}$ inside the particle, and the source of it, which is the induced surface charge, $\sigma = -\boldsymbol{P} \cdot \hat{\boldsymbol{n}}$, as illustrated in Fig.~\ref{fig:flowchart}$(f)$. The polarizability, $\boldsymbol{\alpha} = \boldsymbol{p}/E_{\text{ext}}$, where $\boldsymbol{p}$ is the induced moment given by
\begin{equation} \label{eqn dipole_moment}
  \boldsymbol{p} = \frac{\epsilon_r - 1}{4\pi}\int_V \boldsymbol{E} \mathrm{d}V.
\end{equation}
Considering our problem symmetry, all terms but the dipole one cancel upon integration over volume, leading to: 
\begin{equation} \label{eqn dipole_moment1}
p_x = p_y =0 \quad \mbox{and} \quad p_z = \frac{\epsilon_r - 1}{4\pi} E_z\vert_{r=0} V,
\end{equation}
where $E_z\vert_{r=0}$ is the amplitude of the electric field at the center of the particle. In order to compare with previous works, we introduce the volume normalized polarizability defined by Sihvola et al. \cite{Sihvola01} as: $\overline{\alpha}_j = 4\pi \alpha_j / V$, with $j = x,y,z$. Thus,
\begin{align} \label{eqn E_center}
    \overline{\alpha}_z \approx (\epsilon_r - 1) \frac{E_z\vert_{r=R_\text{min}}}{E_\text{ext}},
\end{align}
because $r=0$ is excluded from the solution domain.

In the following, we fix the parameters $ E_{\text{ext}} = R = b = c = 1$, and let $R_{\text{min}} = 0.01$. In order to keep the dipole contributions on the boundaries of the solution domain of the same order, we take $R_{\text{max}} = 10$. The remaining three parameters: $N$, $a$, and $\epsilon_r$, enable one to assess the emergence of edges and corners, the stretching/squeeze of the geometry, and the dielectric mismatch at the interface.

Let us first assess the present method with dielectric spheroids for which there are analytical solutions and then apply it to quasi-cubes with increasing values of $N$. 

\subsection{Spheroid} \label{sec: application-spheroid}

For $N = 1$, Eq.~(\ref{eqn 4}) defines a unit sphere for $a=1$, and it deforms into a spheroid as $a$ deviates from $a = 1$. In both cases, the induced electric field inside the dielectric particle is uniform and given by an analytical expression \cite{Klimov,Landau_electrodynamics,Jackson,Stratton_book}, thereby providing a benchmark for the accuracy of the proposed ANN approach. Here, we consider a sphere ($a=1$), an oblate shape with $a=2/3$ and a prolate one with $a=3/2$. 

The training process is stopped when the loss function (\ref{eqn loss}) drops below $10^{-5}$. From the electric potential $\phi(r,\theta,\varphi)$, we calculate the Cartesian components of the electric field $E_x$, $E_y$ and $E_z$ by means of the chain rule. Their distributions inside and outside an oblate spheroid with $a=2/3$ and a prolate spheroid with $a=3/2$ with relative dielectric constant $\epsilon_r=6$ are plotted in Fig.~\ref{fig:spheroidal_contour}. Indeed, the calculated induced electric field is quite uniform inside the dielectric particle and decays towards $E_{\text{ext}}$ outside, with values in good agreement with theory \cite{Klimov,Landau_electrodynamics,Stratton_book}.

\begin{figure}
   \centering
   \noindent\makebox[0.75\textwidth]{%
   \includegraphics[width=0.75\textwidth]{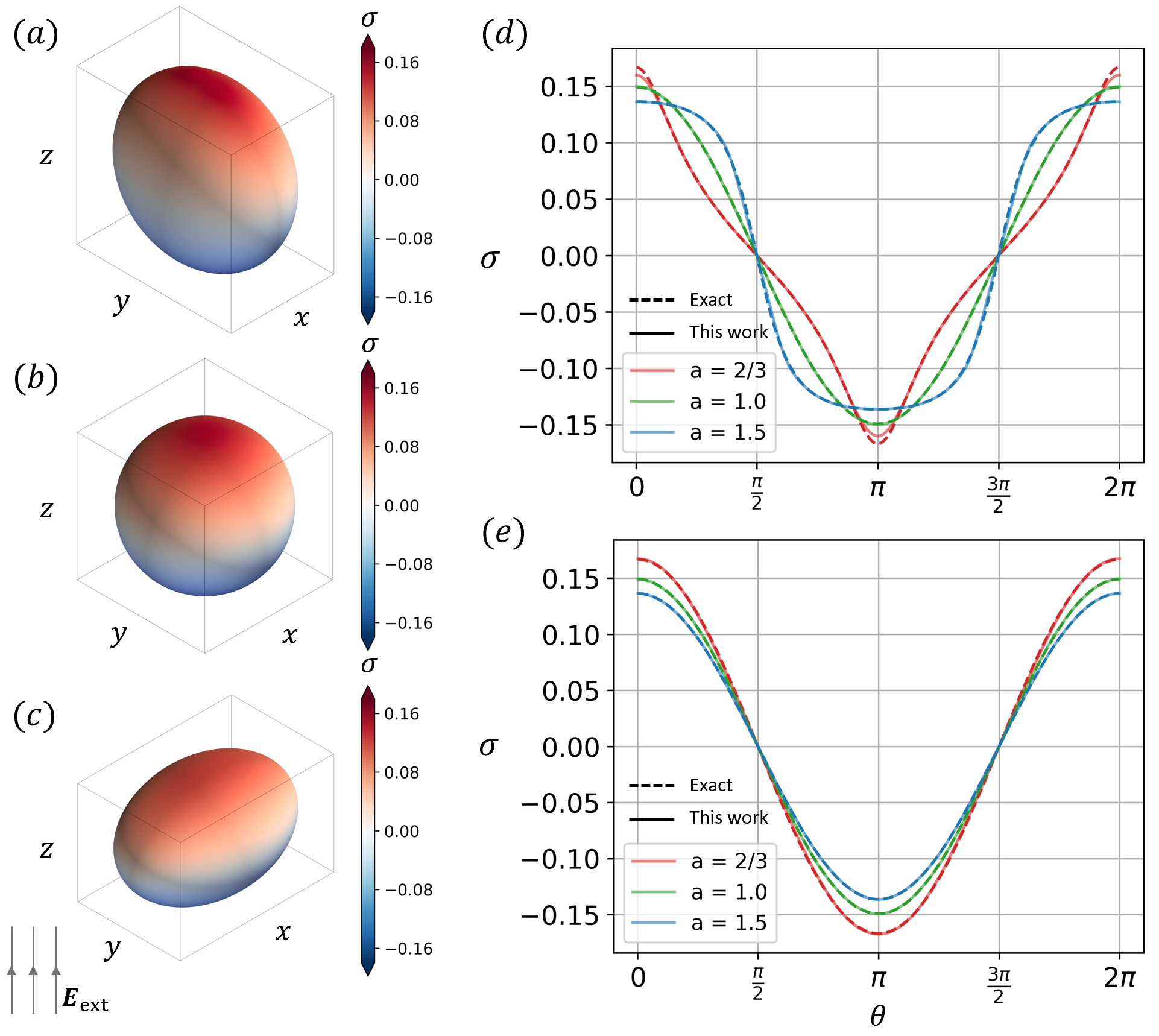}}
   \caption{\label{fig:spheroid_charge} Visualization of surface charge distribution for: $(a)$ an oblate $(a = 2/3, b=c=1)$, $b$ a spherical $(a=1, b=c=1)$, and $(c)$ a prolate $(a = 3/2, b=c=1)$ dielectric particles with $\epsilon_r = 6$. Comparison of the obtained surface charge distributions with their exact solutions on a cross section with $(d)$ $xz$-plane $(\varphi = 0)$ and $(e)$ $yz$-plane $(\varphi = \pi/2)$.}
\end{figure}

In Fig.~\ref{fig:spheroid_charge} we show the distribution of charges on the dielectric interface for spheroidal particles. The accumulation of positive charges on the north pole and negative charges on the south pole leads to an induced electric field which counteracts $\boldsymbol{E}_{\text{ext}}$. Our calculated surface charge distribution agrees nicely with the exact solution except for a small deviation at the tips of the oblate spheroid in Fig.~\ref{fig:spheroid_charge}$(d)$, where the high curvature in the $xz$-plane leads to large variation of $L_{ge}$ components of $\phi_0$, degrading the numerical accuracy.

\begin{table}[!t]
\Huge

\caption{Calculated normalized polarizabilities of dielectric spheroidal particles compared with the corresponding exact values and relative errors.}
\label{table_example}
\centering
\renewcommand{\arraystretch}{1.75}
\resizebox{0.75\textwidth}{!}{
\begin{tabular}{|c|c|c|c|c|c|c|c|c|c|c|c|c|c|}
\hline
\multicolumn{2}{|c|}{\multirow{2}{*}{}} 
& \multicolumn{3}{c|}{$\epsilon_r = 2$} 
& \multicolumn{3}{c|}{$\epsilon_r = 6$}  
& \multicolumn{3}{c|}{$\epsilon_r = 10$}\\ \cline{3-11} 

\multicolumn{2}{|c|}{} & This work & \multicolumn{1}{c|}{Exact} & \multicolumn{1}{l|}{Error(\%)}
& \multicolumn{1}{c|}{This work} & Exact & \multicolumn{1}{c|}{Error(\%)} 
& \multicolumn{1}{c|}{This work} & Exact & \multicolumn{1}{c|}{Error(\%)} \\ \hline

\multirow{3}{*}{\rotatebox[origin=c]{90}{$L \leq 10^{-4}$}}
& Oblate
& 0.7845    & 0.78306  & 0.185       
& 2.1028    & 2.09623  & 0.313  
& 2.5888    & 2.57627  & 0.486    \\ \cline{2-11} 
& Sphere 
& 0.7492  & 0.75000  & 0.107       
& 1.8711  & 1.87500  & 0.208  
& 2.2521  & 2.25000  & 0.093       \\ \cline{2-11} 
& Prolate 
& 0.7245  & 0.72280  & 0.235       
& 1.7179  & 1.71377  & 0.241
& 2.0391  & 2.02175  & 0.858    \\ \hline

\multirow{3}{*}{\rotatebox[origin=c]{90}{$L \leq 10^{-5}$}}
& Oblate 
& 0.78355    & 0.78306  & 0.063       
& 2.09661    & 2.09623  & 0.017   
& 2.57710    & 2.57627  & 0.032     \\ \cline{2-11} 
& Sphere 
& 0.75001   & 0.75000  & 0.001  
& 1.87507   & 1.87500  & 0.004  
& 2.25008   & 2.25000  & 0.004       \\ \cline{2-11} 
& Prolate 
& 0.72314   & 0.72280  & 0.047 
& 1.71379   & 1.71377  & 0.001  
& 2.02157   & 2.02175  & 0.009      \\ \hline
\end{tabular}}

\label{table_spheroid}
\end{table}

To obtain the polarizability, we compute the integral in Eq.~(\ref{eqn dipole_moment}) using a Monte Carlo method \cite{Monte_Carlo16}. The comparison is made for two values reached by the loss function $L$ during optimization. As a reminder, the normalized polarizability (columns ``Exact") of a spheroid with axis ($a, b=c=1$) is: $\overline{\alpha}_z = (\epsilon_r - 1)/ \left[ 1 + (\epsilon_r -1) n_z^\infty \right]$, with $n_z^\infty$ defined in caption of Fig.~\ref{fig:spheroidal_contour}, cf. \cite{Landau_electrodynamics}. Calculated normalized polarizabilities are within less than $1\%$ their exact values depending upon the loss function accuracy, see Table.~\ref{table_spheroid}. More precisely, a $99\%$ accuracy is found for $L \leq 10^{-4}$ and it reaches $99.9\%$ when the loss function is minimized below $L \leq 10^{-5}$. A further minimization to $L \leq 10^{-6}$ would require a significant increase in computation time for a marginal increase in accuracy. Therefore, we shall limit ourselves to a loss level $L \leq 10^{-5}$ for applications to quasi-cubic inclusions in the next section. 

\begin{figure*}
   \centering
   \noindent\makebox[0.95\textwidth]{%
   \includegraphics[width=0.95\textwidth]{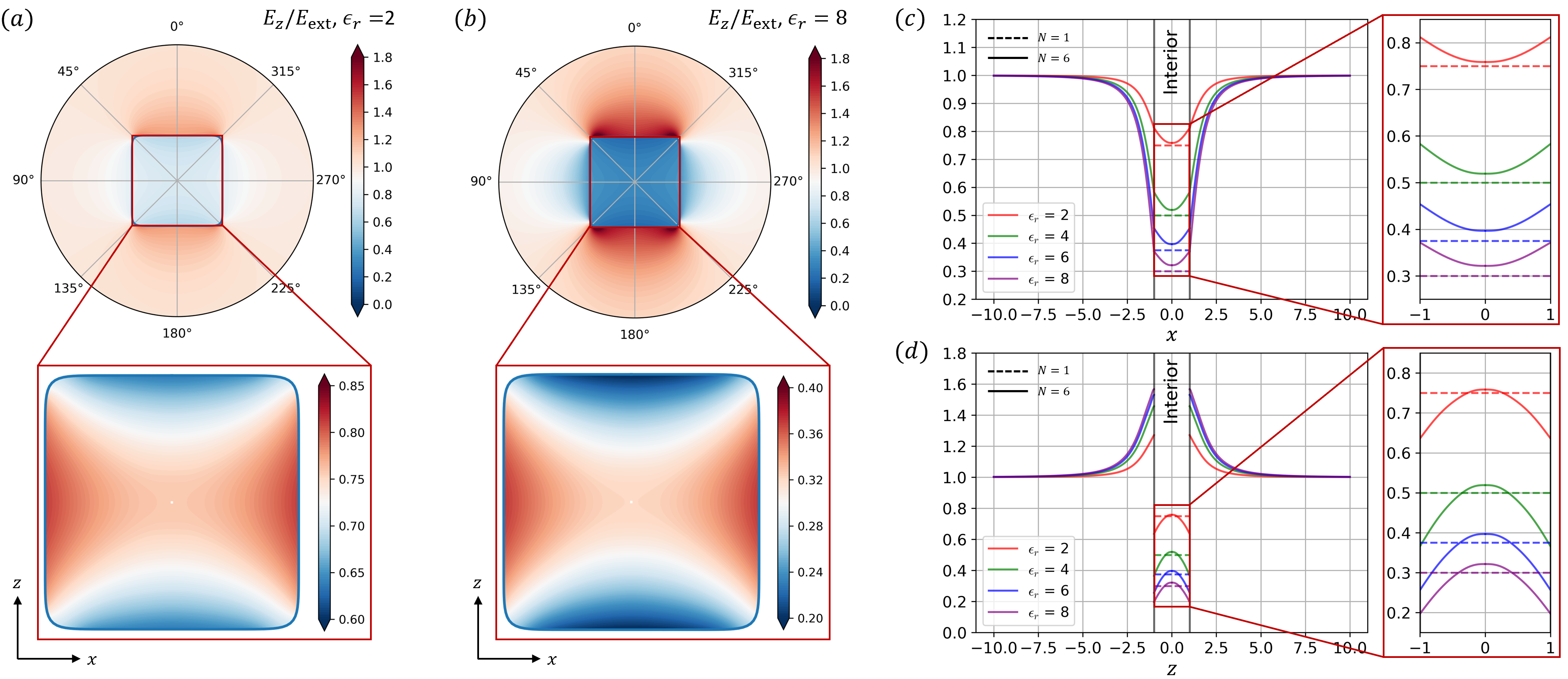}}
   \caption{\label{fig:var_epsilon_r} Contours for the normalized electric field $E_z/E_\text{ext}$ associated with dielectric quasi-cubes ($N = 6$) with $(a)$ $\epsilon_r = 2$ and $(b)$ $\epsilon_r = 8$ on the $xz$-plane. The insets reveal a saddle-shaped field inside particles. Distribution of $E_z/E_\text{ext}$ along $(c)$ $x$- and $(d)$ $z$-axes for dielectric particles with various values of $\epsilon_r = 2,4,6,8$.}
\end{figure*}

\subsection{Quasi-cube} \label{sec: application-quai-cube}

For quasi-cubic particles with $N > 1$, the emergence of edges and corners strongly modifies the electric potential and its derivatives. Fig.~\ref{fig:var_epsilon_r} shows the normalized induced electric field $E_z/E_\text{ext}$ in a quasi-cubic particle with $N=6$ for values of $\epsilon_r = 2,4,6,8$. The strong rise of the electric field along the $z$-axis as one approaches the particle from outside, its discontinuous drop at the interface, and the continuous decrease of $E_z$ along the $x$-axis from infinity to the origin, all together lead to a saddle-shaped field inside the particle, as shown in Fig.~\ref{fig:var_epsilon_r}$(a,b)$. With varying values of $\epsilon_r$, despite an apparent difference of amplitude, the electric field remains qualitatively unchanged. 

\begin{figure}
   \centering
   \noindent\makebox[0.75\textwidth]{%
   \includegraphics[width=0.75\textwidth]{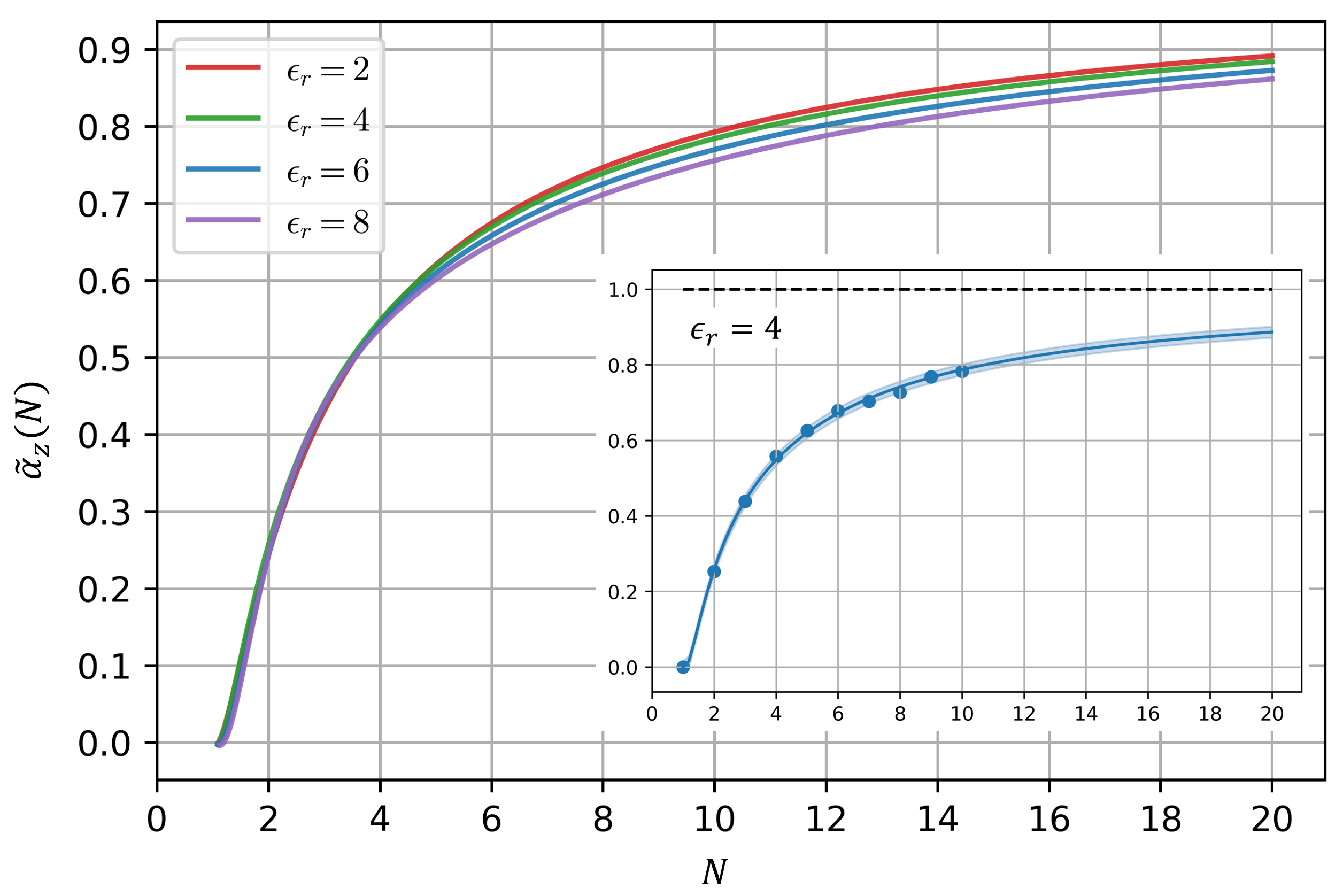}}
   \caption{\label{fig:polarizability} Shape dependent polarizability for quasi-cubic particles with $\epsilon_r = 2, 4, 6, 8$ where, as shown in the inset, points indicate values computed using Eq.~(\ref{feature_scaled_polarizability}) and curves are fitted polynomial functions with an asymptote $\tilde{\alpha}_z\vert_{N \to \infty} = 1$. The shaded regions indicate deviation of the computed data points to the fitted curve. }
\end{figure}

\begin{figure*}
   \centering
   \noindent\makebox[0.95\textwidth]{%
   \includegraphics[width=0.95\textwidth]{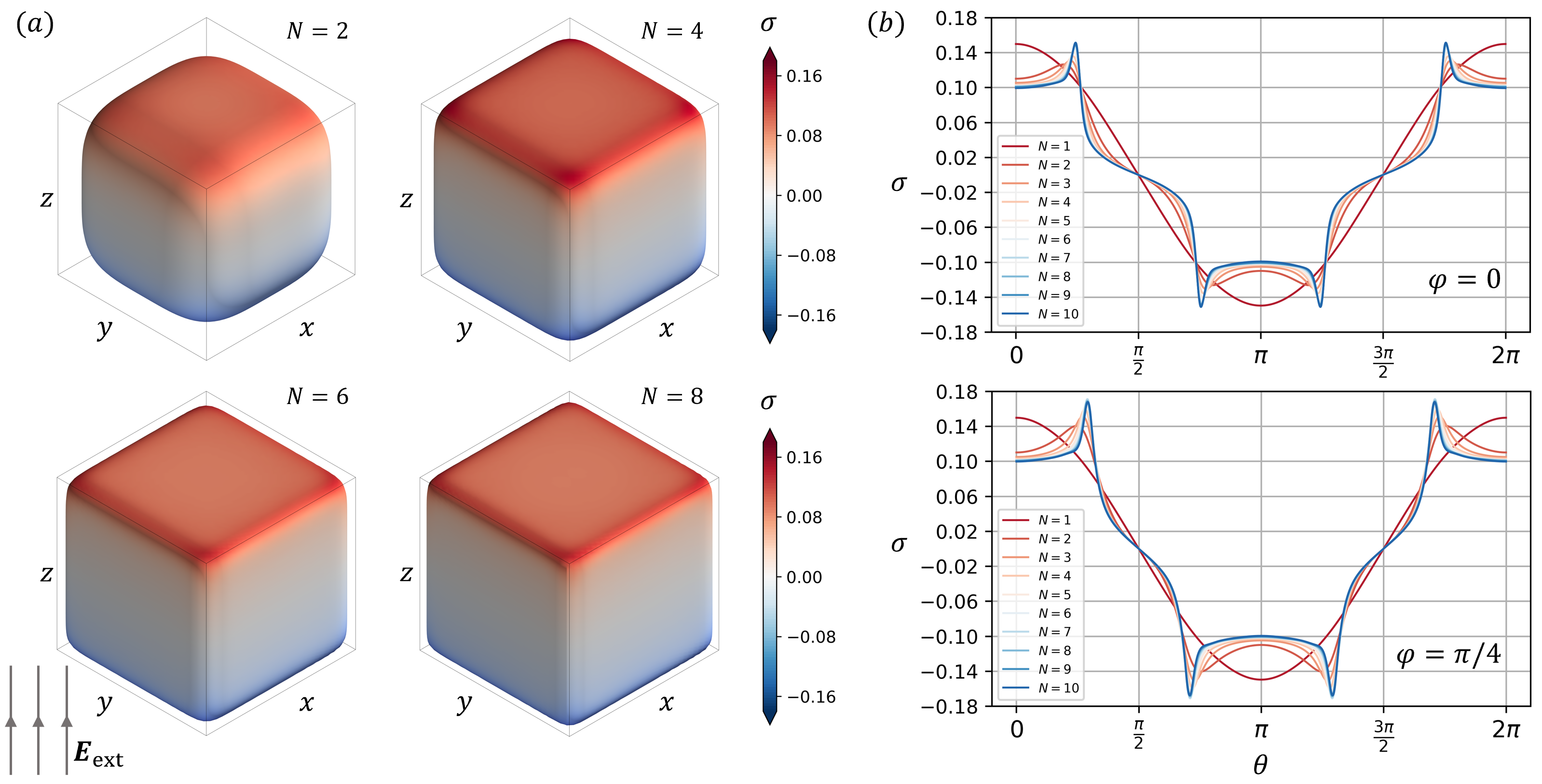}}
   \caption{\label{fig:cubic_charge} $(a)$ Visualization of surface charge on dielectric quasi-cubic particles with $\epsilon_r = 6$ and $N = 2, 4, 6, 8$. $(b)$ Comparison of the obtained surface charge distribution on cross sections $\varphi = 0$ and $\varphi = \pi/4$ of quasi-cubes with $N \in [1, 10]$. }
\end{figure*}

In Fig.~\ref{fig:var_epsilon_r}$(c,d)$, it is observed that the induced electric field $E_z / E_\text{ext}$ at the center of the particle is higher for quasi-cubes than for a sphere, which implies a higher value of the normalized polarizability for quasi-cubes than that for a sphere, cf. Eq.~(\ref{eqn E_center}). To quantify the relation between the polarizability and the shape of the particle, we plot
\begin{align} \label{feature_scaled_polarizability}
    \tilde{\alpha}_z(N) = \frac{\overline{\alpha}_z(N) - \overline{\alpha}_z^1}{ \overline{\alpha}_z^\infty - \overline{\alpha}_z^1},
\end{align}
as a function of $N$ in Fig.~\ref{fig:polarizability}. In most cases, $\overline{\alpha}_z$ obtained using the integral~(\ref{eqn dipole_moment}) is systematically smaller than a direct evaluation using Eq.~(\ref{eqn E_center}) by $0.2\%$. This small deviation can arise from multiple origins, e.g. numerical accuracy, finite size effect, nonzero dipolar component in neural networks, complicating the task of sourcing out the prime contribution. In Eq.~(\ref{feature_scaled_polarizability}), the polarizability of quasi-cubic particles is re-scaled by that of a sphere $\overline{\alpha}^1_z$ and that of a cube $\overline{\alpha}_z^\infty$ to enable a comparison with different values of $\epsilon_r$; hence, $\tilde{\alpha}_z = 0$ for a sphere and $\tilde{\alpha}_z = 1$ for a cube, independent of $\epsilon_r$. For the dependence of $\overline{\alpha}_z^\infty$ upon $\epsilon_r$ we use the approximation formula given in Ref.~\cite{Sihvola01}.  Then, the obtained polarizabilities are fitted to a polynomial function of $N$ which displays an asymptotic behavior $\tilde{\alpha}_z(N) \to 1$ as $N\to\infty$. It is observed that $\tilde{\alpha}_z(N)$ exhibits a rapid transition to almost $80\%$ and $90\%$ of the asymptotic value for $N \leq 10$ and $N \leq 20$, respectively. This numerical result is in accordance with the fact that the geometry of Eq.~(\ref{eqn 4}) converges virtually to a cube for $N > 8$, as seen in Fig.~\ref{fig:super_ellipsoid}. Note that the transition of $\tilde{\alpha}_z (N)$ to its asymptotic value is slower as $\epsilon_r$ increases. 

In order to investigate the behavior of induced surface charges with the emergence of ever sharper edges and corners, we visualize $\sigma$ for quasi-cubic particles with fixed $\epsilon_r = 6$ and increasing values of $N \in [1,10]$ in Fig.~\ref{fig:cubic_charge}. Compared with the sinusoidal charge distribution on the surface of a sphere, charges accumulate towards edges and are peaked at corners as $N$ increases. Such a re-distribution of surface charges leads to an enhanced dipole moment, which in turns leads to the higher value of polarizability observed in Fig.~\ref{fig:polarizability}, thereby a higher electric field at the center of dielectric quasi-cubic particles observed in Fig.~\ref{fig:var_epsilon_r}.

\begin{figure}
\centering
   \noindent\makebox[0.75\textwidth]{%
   \includegraphics[width=0.75\textwidth]{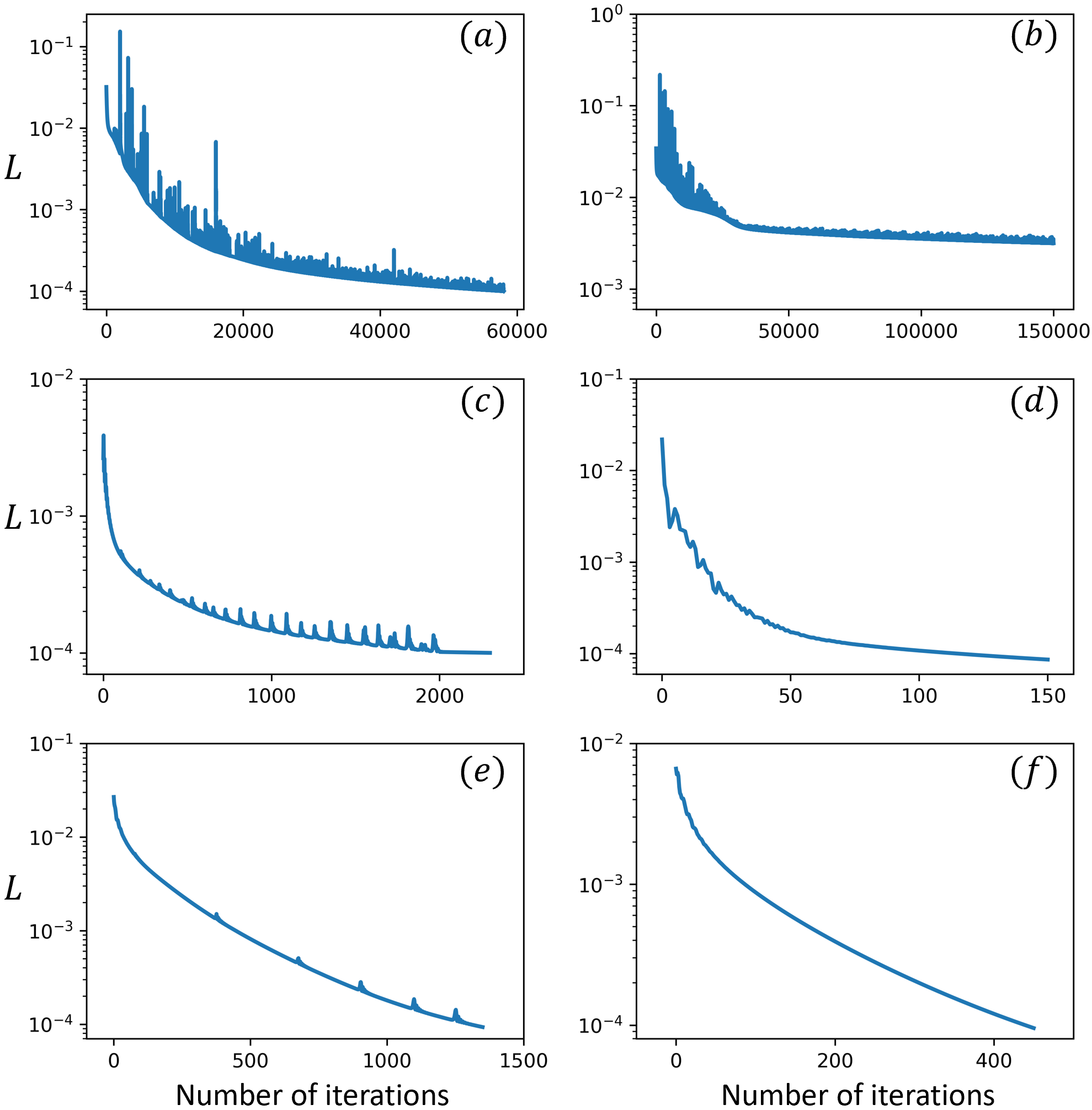}}
   \caption{\label{fig:vanilla} Decay of loss function $L$ for the case $N=3$ and $\epsilon_r=4$ initiated from $(a)$ random conditions using the proposed ansatz (\ref{eqn aphi}); $(b)$ random conditions using the vanilla ansatz; $(c)$ converged solution of $N=2$ and $\epsilon_r = 4$; $(d)$ converged solution of $N=4$ and $\epsilon_r=4$; $(e)$ converged solution of $N=3$ and $\epsilon_r = 2$; $(f)$ converged solution of $N=3$ and $\epsilon_r = 6$. Iteration stops at $L \leq 10^{-4}$ or after $150,000$ iterations. In $(a)$ $L$ decays below $10^{-4}$ within $60,000$ iterations; whereas in $(b)$ $L$ remains of order of $10^{-3}$ after $150,000$ iterations. In $(a)$ and $(b)$, the first $100$ iterations are removed for a better visualization. Due to the slow convergence in $(b)$, the learning rate is decreased by a factor of $1.2$ at every $2000$ iterations until a minimal learning rate $5\times10^{-5}$ is reached.}
\end{figure}

As a baseline comparison, we compare our ansatz with a vanilla ansatz. For the latter, we replaced $H_0$ and $H_1$, cf. Eqs. (\ref{eqn 10}), by multilayer perceptrons with the same network architecture as $\text{NN}_i$ but a linear activation at the output layer. The input variables $r$, $\theta$ and $\varphi$ were rescaled to $[-1,1]$ using the min-max normalization. As explicitly shown in Fig.~{\ref{fig:vanilla}} $(a)$ and $(b)$, the inclusion of physical constraints in ansatz (\ref{eqn 10}) comparatively reduces the optimization effort by a large degree. It is worth noting that in order to capture the transition regime from sphere to cube, classical numerical methods, e.g. finite element, would require a re-mesh and a re-simulation for each different values of $N$. Alternatively, the proposed ANN approach is able to tackle more efficiently this progressive transition. Since each change in $N$ will only affect the boundary conditions and few collocation points near the dielectric interface, while Laplace's equation remains satisfied in the bulk of the solution domain, a minimization initiated from a previous converged solution for, e.g. $N-1$, leads to a drastic reduction in computational time as shown in Fig. \ref{fig:vanilla} $(c)$ and $(d)$. For piece-wise homogeneous dielectric media considered in this work, $\epsilon_r$ appears only through the normal boundary condition (\ref{eqn 2b}). Thus, for simulations with different values of $\epsilon_r$ and a fixed value of $N$, neural networks are committed to learn the change in mismatch, which is induced by a continuous variation of $\epsilon_r$, at the dielectric interface, cf. Fig. \ref{fig:vanilla} $(e)$ and $(f)$. Once the loss function is minimized below the target value, only the converged solutions, which consist of the model variables and the structure of the ANN ansatz, are stored independently of collocation points. Unlike finite element methods whose solutions are defined on meshes, our ANN approach provides a continuous mapping: $(r, \theta, \varphi) \to \phi$ over the entire solution domain $r \in [R_\text{min}, R_\text{max}]$, resembling analytical solutions.

\section{Conclusions} \label{sec: conclusion}

Recent advances in machine learning enable a revisit of longstanding physics problems from a new perspective. We have presented a neural networks based calculation for the response of dielectric particles with shapes varying from sphere to cube, placed in an external uniform electric field. Our ansatz intertwined boundary conditions at the borders of the solution domain, as well as symmetry constraints resulting from the geometry of the particle and the external field, with neural networks. Then, solving Laplace's equation with dielectric boundary conditions was translated to a minimization of a loss function defined in Sec. \ref{sec: loss}. To evaluate the accuracy, we applied the proposed ANN approach to spheroids with various aspect ratios and relative dielectric constants. An overall $99.9\%$ accuracy for the normalized polarizability was achieved, with however, a slight deviation of surface charge from the exact solution at the tips of an oblate spheroid with large curvature, as shown in Fig.~\ref{fig:spheroid_charge}$(d)$. 

As a sphere progressively deforms into a cube, the accumulation of surface charges towards the ever sharper edges and corners leads to a rapid transition of polarizability to its asymptotic value at lower values of $N \leq 10$. This implies that the shape effect has a significant impact on determining the induced polarizability by dielectric nano-particles. The enhanced polarizability with increasing values of $N$ leads to an amplified dipole moment, which, in turn, results in a higher electric field at the center of a quasi-cubic particle than that of a sphere, independent of the relative dielectric constant of the particle.  

Since neural networks are infinitely differentiable, instead of interpolating among discrete values, auto-differentiation enables a conversion of the obtained ANN solution to higher order derivatives, which are again continuous functions. This feature can be advantageous in a broad range of physical applications where higher-order derivatives are of interest. By contrast to the finite element method, where the entire mesh is required for computation, the ANN methods during training use only a fraction of collation points to compute the loss function and update the model parameters by mini-batch stochastic gradient descent techniques. A successive re-sampling enables a complete covering of the solution domain with sufficient number of iterations. Therefore, the mesh-free ANN approach overcomes two major deficiencies of the finite element method stemming from the mesh dependence: (i) the finite differentiability; and (ii) the high memory usage associated with fine mesh.

Finally, it is worth emphasizing some limitations which need to be overcome in future works. Loss functions stemming from physics problems often consist of several components, e.g. governing equation, initial and boundary conditions, symmetries and conservation laws. Since the loss function is not strictly zero, an imbalance in its components can deteriorate the accuracy of the solution. In this work, we introduced weights to balance the amplitudes of each loss component during the training process. However, it is unclear, how does the selection of these weights affect the numerical accuracy. Therefore, it would be beneficial to establish a relation between the accuracy and the relative amplitudes of each components, and devise an adaptive algorithm which automatically adjusts the weights during the training process.

\section*{Appendix} \appendix \label{sec: appendix}

\section{Output activation} \label{sec: output_activation}

\begin{figure}
   \centering
   \noindent\makebox[0.75\textwidth]{%
   \includegraphics[width=0.75\textwidth]{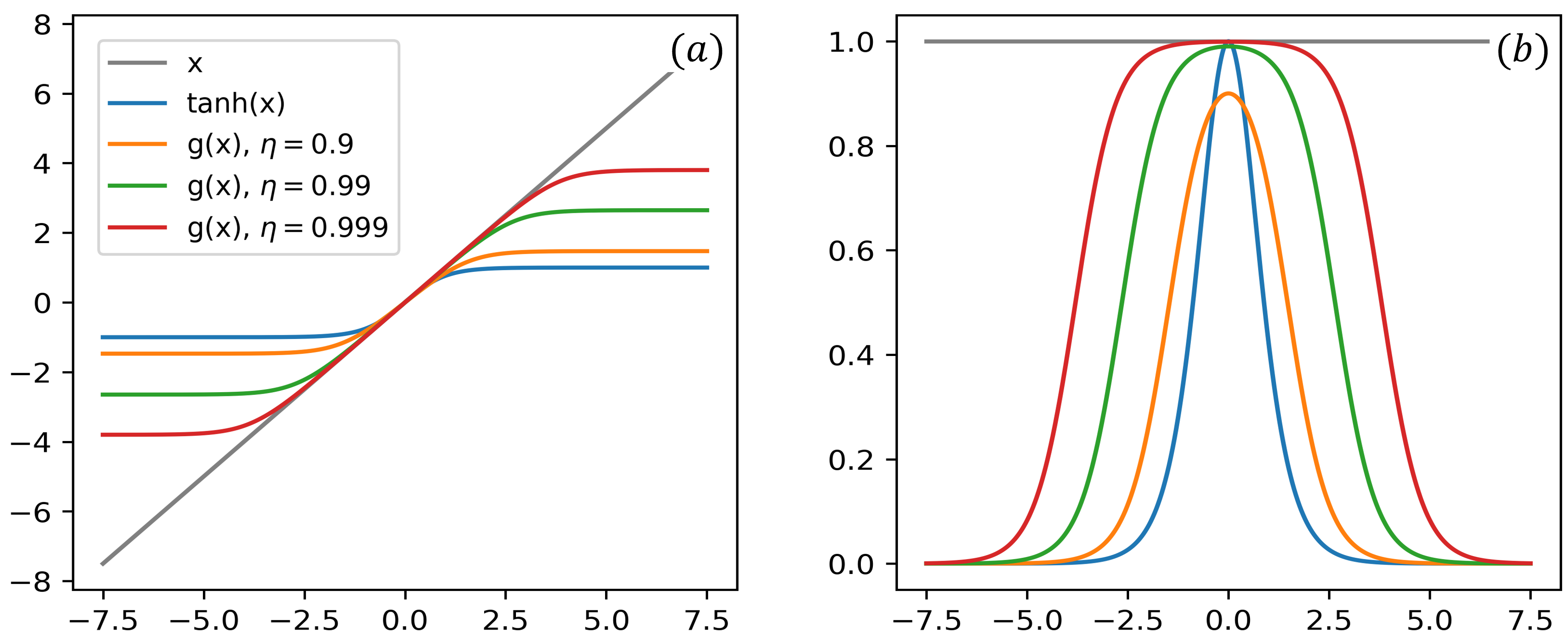}}
   \caption{\label{fig:output_activation} Comparison of $(a)$ linear $x$, $\tanh$, and proposed $g(x)$ activation functions; and $(b)$ their derivatives with respect to $x$. }
\end{figure}

The linear $x$, $\tanh(x)$, and $g(x) = \text{atanh}(\eta \tanh(x))$ activation functions and their derivatives are shown in Fig.~\ref{fig:output_activation}. By varying $\eta \in (0, 1)$, we adjust the output of ANN to an arbitrary bounded interval. Alternatively, one can achieve the same purpose by scaling the $\tanh$ output. However, as observed from Fig.~\ref{fig:output_activation}$(b)$ that a scaling up of $\tanh$ output leads to a sharp variation of gradients. Therefore, we select the activation function at the output layer to be $g(x)$.

\section{Weight tuning}
\begin{figure}
   \centering
   \noindent\makebox[0.75\textwidth]{%
   \includegraphics[width=0.75\textwidth]{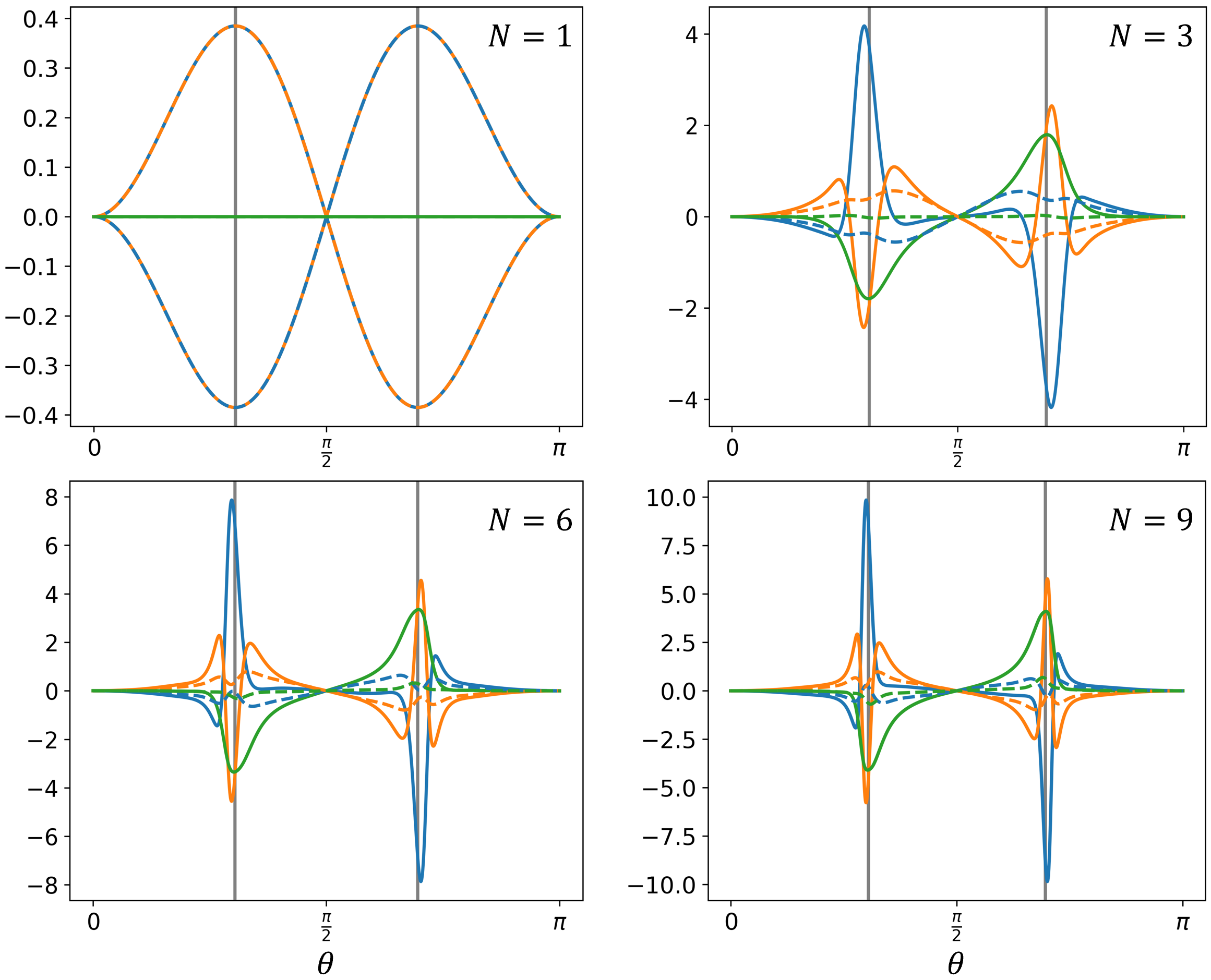}}
   \caption{\label{fig:weight_tuning} Visualization of the radial (blue), the polar (orange) and the azimuthal (green) components of $L_{ge}$ against $\theta$ on the interface with $\varphi=\pi/4$ and $\epsilon_r = 4$. The solid and the dashed lines are associated with $\phi_0$ and $\phi_1$, respectively. The gray vertical lines mark the locations of corners for a canonical cube.}
\end{figure}

The weight $w_{g_0}$ is selected to balance the components of $L_{ge}$ on the dielectric interface. It is observed from Fig.~\ref{fig:weight_tuning} that, due to the emergence of ever sharper corners with increasing $N$, the components of $L_{ge}$ inside and outside of the particle differ in magnitudes substantially. To enable an optimization of $\phi_0$ and $\phi_1$ on the same footing, we take the spherical case $N=1$ as a reference and select $w_{g_0} = 0.3, 0.2, 0.15$ for $N = 3, 6, 9$, respectively. However, since the relative magnitudes of each loss component is not known a priori, an adaptive algorithm which adjusts $w_{g_0}$ during the training can be beneficial.

\section*{Acknowledgment}
The authors would like to thank Steven Blundell and Tan Nguyen for discussions that initiated this work, Liang Mong Koh, Sean Ooi and Kavitha Srinivasan for providing computational resources and Subodh Mhaisalkar for support. The computational work for this article was partially performed on resources of the National Supercomputing Centre, Singapore (https://www.nscc.sg).

\bibliographystyle{unsrt}  


\end{document}